\documentclass[prl,showpacs,footinbib,twocolumn]{revtex4}
\usepackage{graphics,graphicx}

\def\tauphi{$\tau_{\phi}$ }
\def\lphi{$l_{\phi}$ }
\def\gammaphi{$\gamma_{\phi}$ }

\begin{document}
\title{Anomalous temperature dependence of the dephasing time in
mesoscopic Kondo wires}

\author{F\'elicien Schopfer, Christopher B\"auerle, Wilfried 
Rabaud$^{*}$ and Laurent Saminadayar}
\affiliation{Centre de Recherches sur les Tr\`es Basses Temp\'eratures, B.P.
166 X, 38042 Grenoble Cedex 9, France}

\date{\today}


\begin{abstract}
We present measurements of the magnetoconductance of long and narrow
quasi one-dimensional gold wires containing magnetic iron impurities
in a temperature range extending from
$15\,$mK to $4.2\,$K. The dephasing rate extracted from the weak
antilocalisation shows a pronounced plateau in a temperature region of
$300\,$mK\,-\,$800\,$mK, associated with the phase breaking due to the Kondo
effect. Below the Kondo temperature the dephasing rate decreases linearly
with temperature, in contradiction with standard Fermi-liquid theory.
Our data suggest that the formation of a spin glass due to the 
interactions between the magnetic moments are responsible for the observed anomalous
temperature dependence.
\end{abstract}

\maketitle

The understanding of quantum coherence in metallic systems is one of
the most important challenges in mesoscopic physics.  This very
fundamental problem addresses the ``historic" question of the ground
state of an electron gas at zero temperature\cite{pines+nozieres} and 
is also relevant to the recent question of the possible
realisation of quantum computers.  It is well known that the coupling
of electrons to an environment is a very efficient source of
decoherence. In metallic conductors, this coupling may be electron-electron
interaction, electron-phonon interaction or
scattering by magnetic impurities.  In the ``standard'' theory of
metallic conductors\cite{altshuler}, the dephasing rate, defined as
$\gamma_{\phi}=1/\tau_{\phi}$, where $\tau_{\phi}$ is the dephasing
time, is supposed to vanish at zero temperature, as both
electron-electron and electron-phonon interactions go as simple power
laws of the temperature $T$.

Recently, Mohanty et al. have suggested that the phase coherence
time \tauphi saturates at low temperature, in contradiction with
theoretical predictions\cite{mohanty_prl_97}. Based on their own data
on gold wires as well as several experiments of different 
groups, the authors claim that the observed saturation is
universal. However, another experiment on pure silver and pure gold
wires\cite{saclay} shows that the standard theory is
verified in these samples, while other experiments show
that the low temperature behaviour of the dephasing rate depends on
the geometrical parameters of the sample\cite{natelson_prl_00}.
It has been argued that the observed saturation of
\tauphi may be due to inelastic coupling of the electrons to two-level
systems\cite{tls}. On the other hand, an alternative
and controversial interpretation of these experiments is that the observed
saturation is indeed \emph{intrinsic} and due to electron-electron
interactions\cite{zaikin}.

In this context, the understanding of the role of magnetic impurities is
crucial. It is well known that in metals the interaction of conduction
electrons
with magnetic impurities gives rise to the Kondo
effect\cite{kondo}. The most widely studied consequence of this is the logarithmic
increase of the resistivity of metals containing magnetic impurities
below a certain temperature, the Kondo temperature $T_{K}$.
The influence of Kondo impurities on the dephasing rate, on the other hand,
is by far
much less well understood: although the scattering rate and the energy
relaxation time
have been addressed theoretically\cite{nagaoka_suhl}, it should be
emphasized that these models are valid only strictly within the limit
$T>T_{K}$. Below $T_{K}$, the interplay between the magnetic
scattering and the screening of the magnetic impurities by the
surrounding conduction electrons, the Kondo
cloud, is certainly an important source for new physical phenomena. This
points out the crucial need for new experiments
relating the Kondo effect and the dephasing time at low temperatures.

In this Letter we report on the temperature dependence of the
magnetoresistance and resistivity of quasi one-dimensional (1D) long and
narrow Au/Fe Kondo wires down to $15\,$mK. 
We show that the dephasing rate due to Kondo impurities
scattering exhibits a maximum
around $T_{K}$. Below $T_{K}$ 
the dephasing rate continues to decrease, but does not follow the
prediction of standard Fermi-liquid theory.
We show that in this new regime below $T_{K}$, the dephasing rate varies
linearly with temperature and that the interactions between the magnetic
moments are responsible for the
observed anomalous temperature dependence.\begin{table}
\caption{Sample characteristics at $15\,$mK.}
\label{Table}
\begin{tabular}{c c c c c c c c c}
{Sample}&{$w$}&{$t$}&{$L$}&{$R$}&{$D$}&{$l_{e}$}&{$l_{T}$}\\
{}&{(nm)}&{(nm)}&{($\mu$m)}&{($\Omega$)}&
{(cm$^{2}$/s)}&{(nm)}&{($\mu$m)}\\
\hline
\vspace{-2mm} \\
{A}&{$150$}&{$45$}&{$450$}&{$4662$}&{$55.6$}&{$12.0$}&{$1.67$}\\
{B}&{$150$}&{$45$}&{$450$}&{$2236$}&{$115$}&{$24.9$}&{$2.42$}
\end{tabular}
\end{table}

The samples are fabricated on silicon substrate using electron beam
lithography on polymethyl-methacrylate resist. The metal is deposited
using a Joule evaporator and lift-off technique. A $1\,$nm thin titanium layer 
is evaporated prior to the gold evaporation in order to
improve adhesion to the substrate. For the gold evaporation we use two sources
of $99.99\%$ purity with different iron impurity concentrations. Since any
chemical
analysis is impossible on such small samples, we determine the actual iron
impurity
concentration  \textit{via} the resistance increase at low temperature due
to the Kondo effect.
Such a method has the advantage that we directly characterise the purity
of the samples,
which may be quite different from the purity of the sources.
The Au/Fe Kondo system is chosen for the following reasons: besides being
one of the most
extensively studied bulk Kondo
system, it has a Kondo temperature sufficiently low to ensure
that electron dephasing is only weakly affected by electron-phonon
scattering and sufficiently high to be able to perform
measurements down to temperatures well below $T_K$.
Gold as the host metal also
avoids problems with surface oxidation: this has been verified by remeasuring
the samples after 6 months of being exposed to air, and no measurable change
has been observed. The characteristics of the two representative samples
presented here are
given in table \ref{Table}.
The dimensions of the samples are chosen such that they are quasi-1D
with respect to both, the phase-breaking length
$l_\phi$\,=\,$\sqrt{D\tau_{\phi}}$ and the thermal length
$l_T$\,=\,$\sqrt{\hbar D/k_{B}T}$,
$D$ being the diffusion constant, $D$\,=\,1/3\,$v_{F}\,l_{e}$.

\begin{figure}[tbp]
\centerline{\includegraphics*[width=7.0cm]{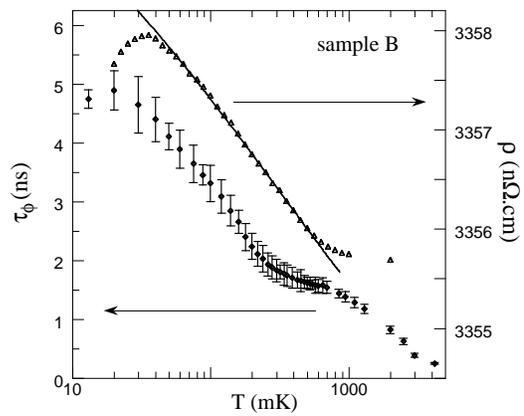}}
\caption{Resistivity and phase coherence time as a function of
temperature. The sample contains $15\,$ppm of Fe impurities.}
\label{tau_phi_R_I_B}
\end{figure}

All the measurements have been carried out using a standard ac
resistance bridge with a current down to 2\,nA for the lowest
temperatures. Extreme care has been taken to filter
external radio-frequency noise: $\pi$-filters are placed at the top of
the cryostat and electrical lines are made of high-loss coaxial
cables\cite{zorin}, leading to an
attenuation of 420\,dB at 20\,GHz. The phase coherence length \lphi is
deduced from magnetoresistance measurements using
standard weak localisation theory\cite{hikami}, where we fix the spin orbit scattering 
length
to 50\,nm, determined at $T\,>\,$1\,K. The coherence time is then obtained
from $\tau_{\phi}=l_{\phi}^{2}/D$.
\begin{figure}[tbp]
\centerline{\includegraphics*[width=7.0cm]{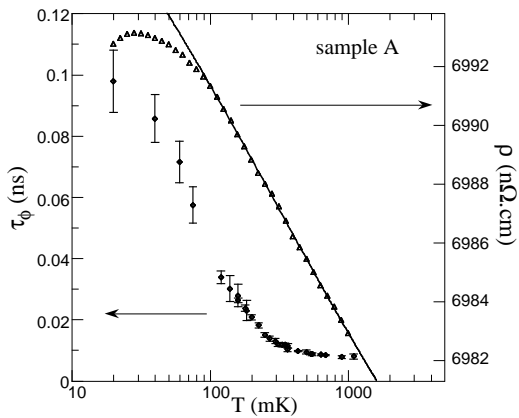}}
\caption{Resistivity and phase coherence time as a function of
temperature. The sample contains $60\,$ppm of Fe impurities.}
\label{tau_phi_R_I_A}
\end{figure}
The temperature dependence of the phase coherence time as well
as the electrical resistivity after subtraction of the electron-electron
contribution for the two samples are displayed in
figures \ref{tau_phi_R_I_A} and \ref{tau_phi_R_I_B}. For the resistivity
measurements, a magnetic field of $200\,$mT is applied to
suppress weak localisation. Both samples
exhibit a logarithmic increase of the resistivity below $1\,$K, a clear
signature of the Kondo effect\cite{chandrasekhar_prl_94}. The temperature dependence of the
electrical resistivity is fitted using Hamann's
law\cite{hamann}: from these fits, we extract a Kondo temperature of $300\,$mK
and an impurity concentration of $60\,$ppm and $15\,$ppm for sample A and
B, respectively\cite{ppm}. At very low temperature, the resistance starts to saturate and
subsequently decreases: this behaviour is associated with the
freezing of the magnetic moments into a spin glass state\cite{laborde_ssc_71}.
To our knowledge, this is the first time that a spin glass transition is
observed in such a diluted mesoscopic Kondo system where weak localisation
is accessible.
The phase coherence time
depends in a similar way on the temperature as the resistivity.  
The plateau observed for \tauphi in the temperature range between 300\,-\,800\,mK is 
associated with the electron
dephasing due to Kondo impurities\cite{bergmann_prl_87,haesendonck_prl_87}.
At temperatures below $T_K$, \tauphi increases again: this is 
expected, as soon as the magnetic impurities become screened by the surrounding conduction
electrons and at low enough temperatures standard Fermi-liquid theory\cite{nozieres} 
should again describe the temperature dependence of \tauphi. 
However, at temperatures well below $T_K$, \tauphi saturates and Fermi-liquid
theory clearly does not describe our experimental data. We rule
out heating effects for the observed
saturation of \tauphi at low temperature for the following reasons:
{\it i)} we have carefully checked the dependence of the phase
coherence time as a function of applied current and no heating effect
has been observed for currents below $8\,$nA; {\it ii)} the downturn
of the resistance at the lowest temperatures, associated with the freezing
of the magnetic impurities into a spin glass, clearly shows that the
electrons are cooled to these low temperatures; {\it
iii)} the temperature of the resistance maximum is consistent with data
from the Au/Fe Kondo system\cite{laborde_ssc_71}.
\begin{figure}[tbp]
\centerline{\includegraphics*[width=6.5cm]{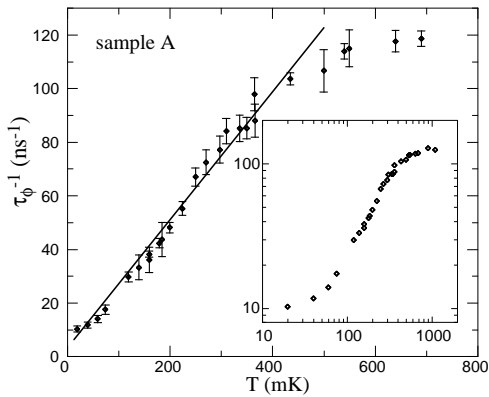}}
\caption{Dephasing rate in the low temperature range: the
straight line is a guide for the eyes. Inset: dephasing rate
over the entire temperature range on a $\log$-$\log$ plot.}
\label{1_tau_phi_A}
\end{figure}

\begin{figure}[tbp]
\centerline{\includegraphics*[width=6.5cm]{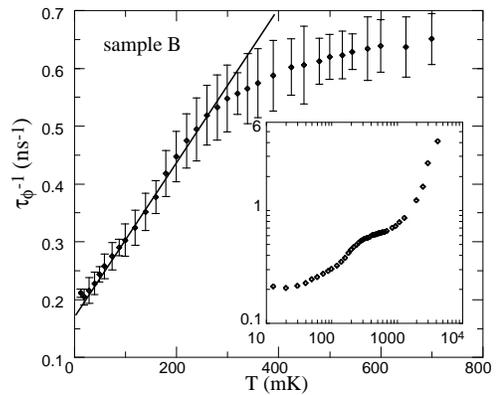}}
\caption{Dephasing rate in the low temperature range: the
straight line is a guide for the eyes. Inset: dephasing rate over the whole
temperature
range on a $\log$-$\log$ plot.}
\label{1_tau_phi_B}
\end{figure}

An explanation of the observed saturation at low temperature of 
$\tau_{\phi}$, however
presently very controversial,
is the possible existence of zero
temperature dephasing due to electron-electron interactions\cite{zaikin}.
Following this model, it is interesting to plot \gammaphi as a function of
temperature on a linear scale
as shown in figures \ref{1_tau_phi_A} and \ref{1_tau_phi_B}: 
\gammaphi varies linearly with temperature
over the entire temperature range below $T_K$ for both samples. In this
respect,
it is tempting to compare our data with the formula
$\gamma_{\phi}=\gamma_{\phi_{0}}+\alpha\, T$ of ref.\cite{zaikin}.
From our measurements we find $\gamma_{\phi_{0}}=0.2\,$ns$^{-1}$ and
$\alpha=0.7\times
10^{-3}\,$ns$^{-1}$\,mK$^{-1}$ for sample B, whereas the theoretical
values, using the geometrical parameters and the experimental 
value of $\gamma_{\phi_{0}}$ of our samples, are $0.7\,$ns$^{-1}$ and
$1.0\times 10^{-3}\,$ns$^{-1}$\,mK$^{-1}$. The agreement found is
reasonable. For sample A, the very high concentration of magnetic
impurities, on the other hand, clearly disallows any comparison with such a
model: in this case, the main decoherence mechanism is due to magnetic
scattering, which is not included in this model. Only the
$\gamma_{\phi_{0}}$ factor is meaningful, and we find an experimental
value of $10\,$ns$^{-1}$ \textit{versus} a theoretical value of $7\,$ns$^{-1}$.

\begin{figure}[tbp]
\centerline{\includegraphics*[width=7.0cm]{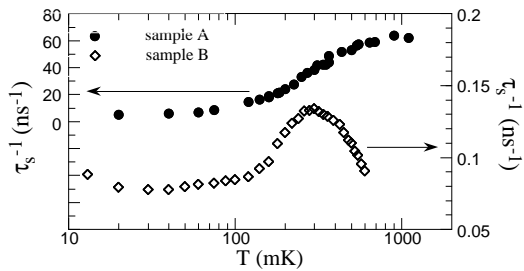}}
\caption{Magnetic scattering rate for sample A and B obtained by subtraction of
the standard dephasing rate from the data of the insets of figures
\ref{1_tau_phi_A}
and \ref{1_tau_phi_B}.}
\label{1_tau_ss}
\end{figure}

It must be stressed that a complete
description of our experimental data has to be done in the framework of the
Kondo effect. To our knowledge, the problem of phase coherence in this
regime at temperatures below $T_K$ has not been addressed
theoretically. Only the unitary limit, where the magnetic impurities are completely 
screened, can be described by Nozi\`{e}res Fermi-liquid 
theory\cite{nozieres}. Such a limit can only be reached
if the impurity concentration is very low; otherwise, a spin-glass
transition will occur and the unitary limit is never reached. 
The crucial question is then: how efficient are these partially screened magnetic impurities for
dephasing? To get an insight into this question, let us try to
separate the magnetic contribution from the total dephasing rate.
The phase breaking rate is given by $\gamma_{\phi}=2\,\gamma_{s}+\gamma_{nm}$,
where $\gamma_s$ is the spin scattering rate and
$\gamma_{nm}$ the non-magnetic scattering rate\cite{haesendonck_prl_87}.
The spin scattering rate can then be obtained by
subtracting \gammaphi measured in samples containing no magnetic impurities.
We assume a temperature dependence for $\gamma_{nm}$ given by
$AT^{2/3}+BT^{3}$, which is justified from measurements on extremely pure gold
wires\cite{saclay}. Coefficient A\,=\,0.8\,(0.6)\,ns$^{-1}$\,K$^{-2/3}$ is 
calculated using the parameters of
sample A (B) and coefficient B\,=\,0.04\,(0.04)\,ns$^{-1}$\,K$^{-3}$ is 
obtained by fitting the data at high
temperatures.
It should be noted that the exact
value of these parameters is not very crucial for the analysis.
In figure \ref{1_tau_ss} we plot the spin scattering rate obtained for
samples A and B.
Both samples exhibit a maximum of $\gamma_s$ around
$T_K$\cite{bergmann_prl_87,haesendonck_prl_87}.
For sample B, $\gamma_s$ decreases with a temperature dependence close
to $T^{1/2}$ below the maximum\cite{bergmann_prl_88,bergmann_prb_89},
whereas for sample A, the maximum is shifted to higher
temperatures and the slope of the subsequent decrease is steeper due to
the higher impurity concentration\cite{bergmann_prb_89}. In neither 
case the $T^{2}$ dependence of $\gamma_s$ of Nozi\`eres 
Fermi-liquid theory\cite{nozieres} nor the 1/ln$^2(T_K$/$T$) dependence for the incomplete 
screened case is observed\cite{glazman}.

At lower temperatures the spin scattering rate saturates for both samples 
and is basically constant down to the lowest temperature. 
It is well known that RKKY interactions between magnetic
impurities lead to the formation of a spin glass.
Spin-spin correlations affect drastically the electronic spin
scattering rate: freezing of magnetic moments into a spin glass state
violates time reversal symmetry, hence leading to a very efficient
dephasing mechanism. 
The change in the temperature dependence of 
$\gamma_s$ can thus be associated with the appearance of spin-spin correlations
between the magnetic impurities. This can also be seen from
the resistivity curve: 
the departure from the logarithmic behaviour appears
approximately at the same temperature as the change in the temperature
dependence of $\gamma_s$. 

In fact, a constant spin scattering rate is
expected for the spin freezing into a
Heisenberg spin glass\cite{bergmann_prb_88}. In this case a reduction of
the scattering rate by a factor
of $S$/($S$+1) compared with the free-magnetic moment system is predicted
\cite{bergmann_prb_88} in good agreement with sample B. 
The ratio obtained from the maximum of $\gamma_s$ and its saturation value, however, varies
strongly with the magnetic impurity concentration. This strong concentration
dependence of both, the reduction of the scattering rate and the saturation
value suggests that the interplay between the Kondo effect and RKKY interactions is 
enhanced in samples with higher impurity concentration.

Our results clearly show that RKKY interactions, associated with the spin glass
freezing, lead to a constant spin
scattering rate and hence yield a finite phase coherence time at very low
temperatures in any system containing even a very small amount of
magnetic impurities.
The understanding of the electron dephasing in the
temperature range below $T_{K}$
is certainly a challenge for theory, but is probably the key point to
interpret properly the experiments carried out on metals as they
often contain magnetic impurities on the ppm level at best, with Kondo
temperatures in the mK range. Measurements at lower
temperature on samples with very low concentration
of impurities, well in the unitary limit, would also be of great interest.
In this case all magnetic impurities are completely screened, and the 
standard Fermi-liquid behaviour should be recovered. This would be
the key test to discriminate between intrinsic dephasing and dephasing
due to Kondo impurities.

In conclusion, we have measured the dephasing rate in quasi-1D gold wires
containing iron impurities down to temperatures of $15\,$mK.
Below the Kondo temperature,
the dephasing rate varies linearly with temperature, clearly in
contradiction with standard Fermi-liquid theory. Our measurements suggest 
that a constant spin scattering rate, associated with the formation of a spin
glass, is responsible for the observed saturation of the phase breaking
rate at low temperature.

We gratefully acknowledge  L. Glazman, A. Zaikin, L.P. 
L\'evy, O. Laborde, J. Souletie, P. Mohanthy, H. Pothier, A. Beno\^\i t 
and F. Hekking for fruitful discussions.
Samples have been made at NanoFab, \textsc{CRTBT-CNRS}. 
Part of this work has been performed at the ``Ultra-Low Temperature
Facility-University of Bayreuth" within a TMR-project of the European
Community (ERBFMGECT-950072). We are indebted to G. Eska, R. K\"onig and I.
Usherov-Marshak for their assistance.


\begin{thebibliography}{99}
\bibitem{ppm}We use formula (4.3) of ref.\cite{hamann}. Note that the 
values for $c$ given in the literature may vary by a factor of two.
\end{thebibliography}
\end{document}